\documentclass{article}
\usepackage{spconf,amsmath,graphicx}
\usepackage{amsmath,graphicx}
\usepackage{amsbsy}
\usepackage{epsfig}
\usepackage{float}
\usepackage{graphicx}
\usepackage{color}
\usepackage{url}
\usepackage{cite}
\usepackage{amssymb}
\usepackage{graphicx}
\usepackage{lipsum}
\usepackage{float}
\usepackage{cuted}
\usepackage{balance}
\usepackage{verbatim}
\usepackage{optidef}

\usepackage{algorithmic}
\usepackage{algorithm2e}
\usepackage{subcaption}
\usepackage[font=small,skip=0pt]{caption}

\newcommand{\beq}{\begin{equation}}
\newcommand{\eeq}{\end{equation}}
\def\bv{\mbox{\boldmath $v$}}

\def\bv{\mbox{\boldmath $v$}}

\def\bC{\mbox{\boldmath $C$}}

\floatstyle{ruled}
\newfloat{algorithm}{tbp}{loa}
\providecommand{\algorithmname}{Algorithm}
\floatname{algorithm}{\protect\algorithmname}

\newcommand{\qedsymbol}{\hspace{\fill}\rule{1.5ex}{1.5ex}}

\makeatother

\newcommand{\FP}[1]{\textcolor{black}{#1}}

\ninept

\title{Goal-oriented communication for edge learning based on the information bottleneck}

\name{Francesco Pezone, Sergio Barbarossa, and Paolo Di Lorenzo\vspace{-.2cm}}
\address{DIET Department, Sapienza University of Rome, Via Eudossiana 18, 00184, Rome, Italy\\ {E-mail: \{francesco.pezone, sergio.barbarossa, paolo.dilorenzo\}@uniroma1.it}\thanks{This work was funded by the H2020 EU-Taiwan Project 5G-CONNI nr. AMD-861459-3 and by MUR under the PRIN Liquid-Edge Project.}\vspace{-0.2cm}}

\begin{document}

\maketitle
\begin{abstract}
Whenever communication takes place to fulfill a goal, an effective way to encode the source data to be transmitted is to use an encoding rule that allows the receiver to meet the requirements of the goal. A formal way to identify the {\it relevant} information with respect to a goal can be obtained exploiting the {\it information bottleneck} (IB) principle. In this paper, we propose a goal-oriented communication system, based on the combination of IB and stochastic optimization. The IB principle is used to design the encoder in order to find an optimal balance between representation complexity and relevance of the encoded data with respect to the goal. Stochastic optimization is then used to adapt the parameters of the IB to find an efficient resource allocation of communication and computation resources. Our goal is to minimize the average energy consumption under constraints on average service delay and accuracy of the learning task applied to the received data in a dynamic scenario. Numerical results assess the performance of the proposed strategy in two cases: regression from Gaussian random variables, where we can exploit closed-form solutions, and image classification using deep neural networks, with adaptive network splitting between transmit and receive sides.
\end{abstract}
\vspace{-.1cm}
\begin{keywords}
Information bottleneck, wireless edge learning, stochastic optimization, resource allocation.
\end{keywords}

\section{Introduction}
\begin{figure}[t]
	\begin{center}
		   \includegraphics[width=8.5 cm]{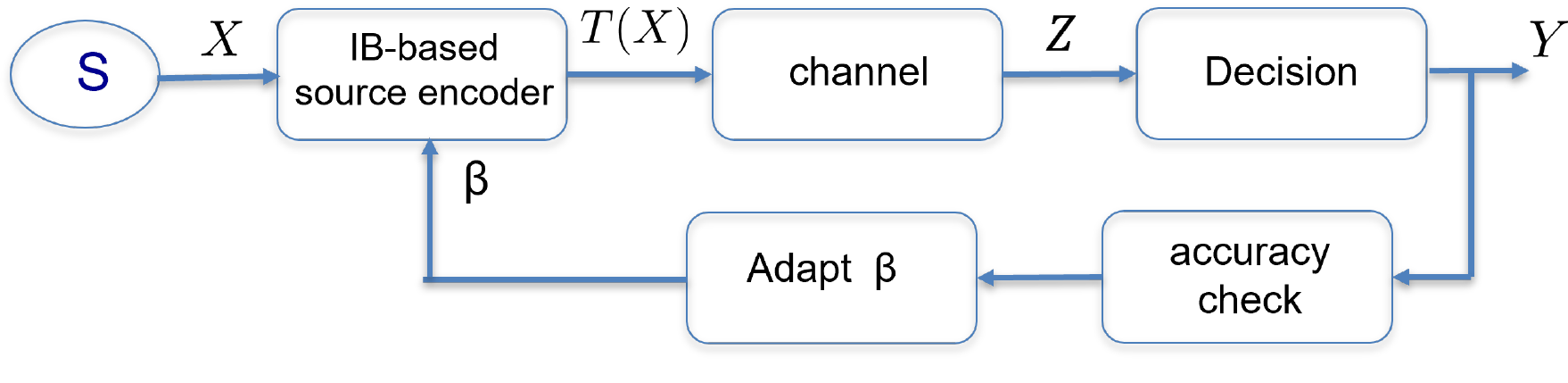}
	\end{center}
	\caption{IB-based goal-oriented communication scheme.}
	\label{IB-scheme}
\end{figure}

Looking at the predictions about the exponential increase of traffic and the associated energy consumption in modern and next generation networks, it is evident that it is time to envisage a new paradigm that should be able to support the expected new services, while limiting the exponential (unsustainable) increase of transmission rate as much as possible. A possible paradigm shift may come from the introduction of semantic communication \cite{strinati20216g}, an idea that can be traced back to Weaver and Shannon himself. While Shannon in his work deliberately focused only on the technical level, it is maybe time to move to the semantic level and be concerned about the recovery of semantics, or meaning, underlying the sequence of transmitted symbols. In this work, we propose an approach that improves the efficiency of the whole transmission system, applicable whenever communication takes place to fulfill a goal. In such a case, it is the goal that assigns a meaning to the communication. An important example comes from the introduction of machine learning (ML) methods to extract information from data collected by a set of sensors and sent to a fusion center (FC) for processing. In this case, the goal is to achieve a sufficient level of accuracy in the decision taken by the FC, and {\it not} the recovery of all the transmitted symbols. The situation can be explained by referring to Fig. \ref{IB-scheme}, where $X$ is a random variable modeling the observation, possibly resulting from a generative (probabilistic) model that associates $X$ to a label $Y$. The goal of the receiver is to recover an estimation $\hat{Y}$ of $Y$ with a sufficient level of accuracy. The idea is to encode the source data $X$ in order to send only the {\it relevant} information necessary to recover the variable $Y$ at the receiver, but non necessarily $X$.

The goal-oriented communication paradigm considered in this work falls into the context of \textit{wireless edge machine learning} \cite{park2019wireless,chen2019deep,Merl2021EML,skatchkovsky2019optimizing,amiri2020machine,wang2018edge,mohammad2019adaptive}, where the inference process typically requires not only high learning accuracy and reliability, but also a very short response time necessary for autonomous decision making in highly dynamic wireless environments. The challenge of edge ML is then to jointly optimize inference, training, communication, computation, and control under end-to-end latency, reliability, and learning performance requirements. For instance, in \cite{Merl2021EML}, the trade-off between energy expenditure, latency, and accuracy of a learning task was explored by properly adapting the number of bits used to quantize the data to be transmitted. In this paper, we also aim to act on the source encoder of edge devices, but hinging on the information bottleneck (IB) principle \cite{tishby99}, which amounts to finding the encoding rule $T(X)$ that is {\it maximally informative} about $Y$, while minimizing the {\it complexity} associated with the representation of $X$. In formulas, the encoding rule is given by the probabilistic mapping, given by the conditional probability $p_{T/X}(t/x)$, that solves the IB problem:
\begin{equation}
    \min_{p_{T/X}(t|x)} \;\; I(X; T) - \beta \cdot I(T; Y),
    \label{eq:IB main problem}
\end{equation}
where $I(X; Y)$ indicates the mutual information between $X$ and $Y$.
The two terms appearing in the objective function are the {\it relevance} $I(T; Y)$ of $T$ with respect to $Y$ and the {\it complexity} of $T$ in representing $X$; $\beta$ is a non-negative parameter that allows us to explore the trade-off between relevance and complexity. Recent excellent surveys on the IB principle and its application to learning are \cite{goldfeld2020information} and \cite{zaidi2020information}. The IB principle is closely related to the concept of minimal sufficient statistics (MSS), the difference being that the IB encoding rule is probabilistic, while the MSS is deterministic; furthermore, the IB allows us to explore the trade-off between relevance and complexity. The IB method is also closely related to  Remote Source Coding (RSC) \cite{wolf1970transmission} and to canonical correlation analysis (CCA) \cite{hotelling1992relations}. 

\textit{Contribution of the paper:} In this paper, we propose a novel goal-oriented communication scheme for edge learning, as depicted in Fig. \ref{IB-scheme}, which exploits the IB principle to limit the transmission rate to the only information that is relevant for the inference task that takes place at the destination, and then merges the IB method with stochastic optimization in order to adapt the complexity/relevance trade-off parameter $\beta$ so as to approach the optimal trade-off between energy consumption, service delay and inference accuracy. 
\textcolor{black}{The idea of exploiting the IB principle in goal-oriented communication was initially suggested in \cite{strinati20216g} and recently analyzed in more depth in \cite{shao2021learning}, using the variational IB to extend the applicability of the IB principle and using a variable-length feature encoding, adjusted to the channel conditions}. Differently from \cite{shao2021learning}, we exploit the IB  principle to reach an optimal balance between the three major performance parameters of an edge learning system: energy consumption, service delay and inference accuracy. Furthermore, we consider a multi-user system where a set of devices send their data to a single edge server that handles the different tasks by optimizing the percentage of CPU time allocated to each requesting device. The proposed framework is then applied to two specific learning tasks: Regression from Gaussian random variables, and image classification using a deep neural network. Numerical results illustrate the performance of the proposed  goal-oriented communication system.

\section{Information bottleneck principle}
In this section, we will review some basic properties of the IB principle, as relevant for the rest of this paper. The solution of the IB problem, as given in \eqref{eq:IB main problem} is not easy, as the problem is nonconvex. Nevertheless, for discrete random variables, the problem can be solved using an iterative algorithm with provable convergence guarantees \cite{tishby99}. The solution for continuous random variables is much harder to find, with the noticeable exception  occurring when $\mathbf{x}$ and $\mathbf{y}$ are jointly Gaussian random vectors. Let us consider the case in which $\mathbf{x}\sim {\cal N}(\mathbf{0}, \mathbf{C}_{X})$ and $\mathbf{y}\sim {\cal N}(\mathbf{0}, \mathbf{C}_{Y})$ are centered multivariate jointly Gaussian vectors of dimension $d_x$ and $d_y$, respectively. Let us also denote by $\mathbf{C}_{XY}$ the cross-covariance between $\mathbf{x}$ and $\mathbf{y}$. In such a case, the solution is known in closed form \cite{GIB}. In particular, the boundary of the relevance/complexity region can be explored using a linear encoder
$T(\mathbf{x})  = \mathbf{A} \mathbf{x} + \xi$, where $\xi \sim  \mathcal{N}(0, \Sigma_{\xi})$ is a Gaussian noise statistically independent of $(\mathbf{x}, \mathbf{y})$. For any given value of the trade-off parameter $\beta$, the optimal matrix $\mathbf{A}$ assumes the following structure: 
\begin{equation}
\mathbf{A} = \left\{\begin{matrix}
            [\mathbf{0}^T;...;\mathbf{0}^T] & 0 \leq \beta \leq \beta_1^c\\
            [\alpha_1\mathbf{v}_1^T; \mathbf{0}^T;...;\mathbf{0}^T] & \beta_1^c \leq \beta \leq \beta_2^c\\
            [\alpha_1 \mathbf{v}_1^T;\alpha_2\mathbf{v}_2^T;\mathbf{0}^T;\ldots;\mathbf{0}^T] & \beta_2^c \leq \beta \leq \beta_3^c\\
            \vdots\\
            [\alpha_1 \mathbf{v}_1^T;\alpha_2\mathbf{v}_2^T;\ldots;\alpha_N\mathbf{v}_N^T] & \beta_{N-1}^c \leq \beta \leq \beta_N^c
            \end{matrix}\right.
\label{eq:Matrice A}
\end{equation}
where $\bv_i$'s are the left eigenvectors of matrix $\bC_{X/Y}\,\bC_X^{-1}$, sorted by their corresponding ascending eigenvalues $\lambda_i$, for all $i=1, \ldots, N$; also, $\beta_i^c = \frac{1}{1-\lambda_i}$ denote the critical values of $\beta$, $\alpha_i=\sqrt{\frac{\beta(1-\lambda_i)-1}{\lambda_i r_i}}$, with $r_i = \mathbf{v}_i^T\mathbf{C}_{X} \mathbf{v}_i$,  for all $i=1, \ldots, N$. The eigenvectors $\{\mathbf{v}_i\}_{i=1}^N$ coincide with the canonical correlation analysis (CCA) vectors \cite{hotelling1992relations}. The structure in \eqref{eq:Matrice A} makes clear the effect of the IB-encoder: When $\beta$ is very small, few data are transmitted, because more importance is given to the complexity of the representation; conversely, as $\beta$ increases, more and more eigenvectors are added (thus increasing the rank of $\mathbf{A}$), since more importance is given to the relevance of the data to be transmitted in order to facilitate the recovery of $\mathbf{y}$. In the Gaussian case, for any given $\beta$, it is also possible to write in closed form the mutual information between the pairs $(X, T_\beta)$ and $(T_{\beta}, Y))$ as \cite{GIB}:
\begin{align}
    &I(X; T_{\beta})=\frac{1}{2} \sum_{i=1}^{n_{\beta}} \log_2\left((\beta-1)\frac{1-\lambda_i}{\lambda_i}\right)\label{I(X;T)}\\
    &I(T_{\beta}; Y)=I(X; T_{\beta})-\frac{1}{2} \sum_{i=1}^{n_{\beta}}\log_2\left(\beta(1-\lambda_i)\right), \label{I(T;Y)}
\end{align}
where $n_{\beta}$ is the maximal index $i$ such that $\beta\ge 1/(1-\lambda_i)$. Interestingly, even though the dimensionality of $T_\beta$ changes discontinuosly with $\beta$ (with the discontinuities represented by the critical values $\beta_{i}^c$), the curve $(I(X; T_{\beta}), I(T_{\beta}; Y))$ changes continuosly with $\beta$. Finally, in the non-Gaussian case, a closed form solution is not known. However, an IB-based encoder can still be found by defining a {\it variational} (lower) bound of the IB-Lagrangian \eqref{eq:IB main problem}, which can be optimized more easily than the IB-Lagrangian directly \cite{zaidi2020information}. 

\section{Dynamic Edge Learning based on the Information Bottleneck Principle}
In this section we propose a dynamic resource allocation strategy for the scheme depicted in Fig. \ref{IB-scheme}. The goal of the proposed method is the minimization of the average energy consumption, under constraints on the average service delay and the average accuracy of the learning task. We consider a scenario composed of $K$ devices sending data to an edge server using an IB-based encoder. The resources to be dynamically allocated include computation resources, namely the CPU clock rates used at the mobile devices and at the server, and communication resources, e.g., the transmission rates and the trade-off parameters $\beta_k$ used in each IB-based encoder. The time axis is slotted in intervals indexed by $t$ and the allocation strategy is be dynamic. The models used for power consumption, delay, and learning accuracy of the edge learning task are described below.\\ 

\subsection{Power Consumption}
We consider three sources of power consumption due to processing at the devices and at the server sides, and communication between devices and server. In particular, denoting by $f_k^d(t) $ the clock frequency of the CPU of the device $k$, the power spent by device $k$ to carry out the computations to obtain the transformation $T_k(X_k)$ is:
\begin{equation}
    p_k^p(t) = \eta_k (f_k^d(t))^3,
\label{eq:Power Device freq GIB}
\end{equation}
where $ \eta_k $ is the effective switched capacitance of processor $k$ \cite{burd1996processors}. 

The wireless channel from each device to the edge server is characterized by a bandwidth $B_k$ and a flat-fading coefficient $\textnormal{h}_k$; the noise power spectral density at the receiver is $N_0$.
Denoting by $R_k(t)$ the data rate (bit/sec) used in slot $t$ by device $k$, the relation between the transmit power $p_k^t(t)$ and the rate can be expressed using Shannon's formula:
\begin{equation}
    p_k^t(t) =  \frac{B_k N_0}{\textnormal{h}_k(t)} \left[  {\rm exp} \left(\frac{R_k(t) ln(2)}{B_k} \right)   -1 \right].
\label{eq: Power antenna}
\end{equation}
On the server side, denoting by $ f_c (t) $ its CPU clock rate, with $f_c (t)\in [0, f_{max}]$, the power spent for computing is:
\begin{equation}
    P^s(t) = \eta f_c^3(t)
\label{eq:Energy ES GIB}
\end{equation}
where $ \eta $ is the effective switched capacitance of the server processor. The total power spent by the system at time $t$ is then given by: 
\begin{equation}
    P_{tot}(t) = \sum_{k=1}^K \left[p_k^t(t) + p_k^p(t)\right] + P^s(t).
\label{eq:Energy totale GIB}
\end{equation}
\subsection{Edge Learning Delay}
Let us consider now the delays associated to computation and communication. Letting $ C_k^d (t) $ be the number of operations needed by the IB-based encoder, the corresponding computation delay is:
\begin{equation}
    L_k^p(t) =\frac{C_k^d(t)}{f_k^d(t)}.  
    \label{eq:Delay device processing GIB}
\end{equation}
Let us now model the communication delay. Denoting by $\beta_k(t)$ the trade-off parameter used at time slot $t$ for device $k$, and by $T_{\beta_k}(t)$ the corresponding relevance value, the number of bits used to encode $T_{\beta_k}(t)$ can be computed as $h(T_{\beta_k}(t))$, with $h(\cdot)$ representing the differential entropy, and it depends on the IB trade-off parameter $\beta_k(t)$ at time $t$. In the Gaussian case, its value is given by \eqref{I(T;Y)}. Now, denoting by $R_k(t)$ the transmission rate used to send $T_{\beta_k}(t)$ to the server during slot $t$, the corresponding transmission delay behaves as follows:
\begin{equation}
    L_k^t(t) = \frac{h(T_{\beta_k}(t))}{R_k(t)}. 
\label{eq:Delay device trasmissione GIB}
\end{equation}
Once the data are received by the server, there is an additional delay associated to estimating $Y_k$ from the received data. We assume that the server assigns a portion $T_k/T$ of its computing time, or equivalently  a  portion $f_k (t)$ of its clock rate $f_c (t)$, to each device,  with
    $\sum_{k=1}^K f_k(t) \leq f_c(t)$.
The amount of computations at the server needed to compute $Y_k$ depends on the size of the output $Y_k$, which is fixed, and on the size of  $T_{\beta_k}(t)$,  which depends on $ \beta_k$. Thus, let $C_{\beta_k}^s$ be the number of CPU cycles needed to perform the computation of $Y_k$. The overall processing delay at the server is then given by:
\begin{equation}
    L_k^s(t) = \frac{C_{\beta_k}^s}{f_k(t)}.  \label{eq:Delay ES GIB}
\end{equation}
Finally, the overall (computation plus communication) delay occurring in slot $t$ for each device $k$ writes as:
\begin{equation}
    L_k^{tot}(t) = L_k^p(t) + L_k^t(t) + L_k^s(t), \quad k=1,\ldots,K.
\label{eq:Delay totale GIB}
\end{equation}
\subsection{Learning Accuracy}
In this paragraph, we assess the performance achievable by the proposed method in recovering the decision variable $Y_k$ from the encoded data using the Mean Square Error (MSE) criterion. 
In the Gaussian case, denoting by $\mathbf{\Sigma}_{Y_k}$ and $\mathbf{\Sigma}_{T_{\beta_k}}$ the covariance matrices of $Y_k$ and $T_{\beta_k}$ and with $\mathbf{\Sigma}_{Y_k T_{\beta_k}}$ the cross-covariance between $Y_k$ and $T_{\beta_k}$ the MSE can be written in closed form as:
\begin{equation}
\begin{aligned}
   MSE_{\beta_k}(Y_k,\hat{Y}_k) 
   = {\rm tr}(\mathbf{\Sigma}_{Y_k}) - {\rm tr}\left(\mathbf{\Sigma}_{Y_k T_{\beta_k}} \mathbf{\Sigma}_{T_{\beta_k}}^{-1}\mathbf{\Sigma}_{Y_k T_{\beta_k}}^T\right).
\label{eq:MSE GIB}
\end{aligned}
\end{equation}
As expected, the $MSE$ in (\ref{eq:MSE GIB}) depends on $ \beta_k $; typically, if $ \beta_k $ increases, the second term increases as well and therefore the error decreases. Normalizing the $MSE$ to $tr(\Sigma_{Y_k})$, we define the accuracy metric as $G(\beta_k) = NMSE_{\beta_k}(Y_k,\hat{Y}_k)$.

\subsection{Problem formulation}
The optimization problem can then be cast as follows:
\begin{align}\label{eq:initial_opt_problem}
\min_{{\Phi}(t)}&{\;\;\displaystyle\lim_{T \to +\infty}\; \frac{1}{T} \sum_{t=1}^T  \mathbb{E}[P_{tot}(t)] }\nonumber\\
&\hbox{s.t.} \;\;{\displaystyle\lim_{T \to +\infty}\; \frac{1}{T} \sum_{t=1}^T  \mathbb{E}[L_k^{tot}(t)] \leq L_k^{avg}\quad \forall k }\nonumber\\
&\quad\;\;{\displaystyle \lim_{T \to +\infty}\; \frac{1}{T} \sum_{t=1}^T  \mathbb{E}[G_k(t)] \leq G_k^{avg}\quad \forall k }\\
&\quad\;\;{0 \leq f_k^d(t) \leq f_{k, max}^d \quad \forall k,t }\nonumber\\
&\quad\;\;{0 \leq R_k(t) \leq R_{k, max}(t) \quad \forall k,t }\nonumber\\
&\quad\;\;{\beta_k(t) \in \mathcal{B}_k \quad \forall k,t}\nonumber\\
&\quad\;\;{0 \leq f_c(t) \leq f_{max} \quad \forall t }\nonumber\\
&\quad\;\;{f_k(t) \geq 0 \quad \forall k,t}, \qquad {\sum_{k=1}^K f_k(t) \leq f_c(t)  \quad \forall t},\nonumber
\end{align}
where ${\Phi}(t) = [\{f_k^d(t)\}_k,\{ R_k(t)\}_k, \{\beta_k(t)\}_k, \{f_k(t)\}_k, f_c(t)]$ is the vector of the optimization variables at time $t$. The expected values are computed with respect to the channel coefficients and the arrival rates of the computing tasks. The constraints of \eqref{eq:initial_opt_problem} have the following meaning: (i) the average latency cannot exceed a predefined value $L_k^{avg}$; (ii) the average performance metric cannot overcome a predefined value $G_k^{avg}$; (iii) the other constraints impose instantaneous bounds on the resource variables. This problem is complex, since we do not have access to the statistics of the involved random variables. In the next section, we show how to handle it  resorting to stochastic optimization \cite{Neely10}.

\subsection{Dynamic Resource Allocation via Stochastic Optimization}
The first step to handle problem \eqref{eq:initial_opt_problem}  is to introduce two \textit{virtual queues} for each device, associated to the long-term delay and accuracy constraints, respectively.  Proceeding as in \cite{Merl2021EML}, these two virtual queues evolve as follows:
\begin{align}
  Z_k(t+1) &= \max [0,Z_k(t) + \varepsilon_k(L_k^{tot}(t) - Q_k^{avg})], \label{eq:Z_k}\\
  S_k(t+1) &= \max [0,S_k(t) + \nu_k(G_k(t) - G_k^{avg})],  \label{eq:Y_k}
\end{align} 
$k=1,\ldots,K$, where $\varepsilon_k$ and $\nu_k $ are positive step-sizes. The goal is to satisfy the constraints on the average values by enforcing the stability of the associates virtual queues \cite{Neely10}. To this aim, we define the Lyapunov function 
    $L({\Theta}(t)) = \frac{1}{2} \sum_{k=1}^K Z_k^2(t) + S_k^2(t)$, 
where ${\Theta}(t)  = [ \{Z_k(t)\}_k , \{S_k(t)\}_k] $. Then, we introduce the \textit{drift-plus-penalty} function: 
\begin{equation}
\begin{aligned}
    \Delta_p(\Theta(t)) &= \mathbb{E}\left[L({\Theta}(t+1))-L({\Theta}(t))+V\cdot P_{tot}(t)  \;\Big|\; \Theta(t)\right],
\label{eq:drift-penalty}
\end{aligned}
\end{equation}
whose minimization aims to stabilize the virtual queues in (\ref{eq:Z_k})-(\ref{eq:Y_k}), while promoting low-power solutions for large values of the parameter $V$. Using stochastic approximation arguments \cite{Neely10}, we remove the expectation per each time-slot $t$ and minimize a suitable upper-bound of \eqref{eq:drift-penalty}, thus leading to the following per-slot deterministic optimization problem:

\begin{align}\label{eq:deterministic_problem}
\min_{{\Phi}(t)}&{\;\;\sum_{k=1}^K \bigg[ \varepsilon_k Z_k(t)L_k^{tot}(t) + \nu_k S_k(t) G_k(\beta_k)\bigg] + V P_{tot}(t)}\nonumber\\
& \hbox{s.t.} \;\;{\Phi(t) \in \mathcal{Z}(t) }\nonumber
\end{align}
where $\mathcal{Z}(t)$ indicates the space of possible solutions given by the constraints on the optimization variables.

This deterministic per-slot optimization, to be solved for each time-slot, can be decoupled into two sub-problems, one associated with the device parameters, i.e., $[\{f_k^d(t)\}_k,\{ R_k(t)\}_k,\{\beta_k(t)\}_k$, and the other associated with the edge server parameters, i.e., $[\{f_k(t)\}_k, f_c(t)]$. Interestingly, both sub-problems admit simple closed form solutions (derivations are omitted due to lack of space). In particular, for a fixed value $\beta_k(t) \in \mathcal{B}_k$, the optimal rate and CPU frequency of device $k$ at time $t$ are given by:  
\begin{equation}
    R_k^*(t) = \frac{2 B_k}{ln(2)}\; W\! \!\left(\sqrt{\frac{Z_k(t)\; h(T_{\beta_k}(t))\; ln(2)\; \textnormal{h}_k(t)}{4 B_k^2\;V \;N_0}}\right)\; \Biggr|_0^{R_{k, max}(t)}
\label{eq:Optim_rate}
\end{equation}
\begin{equation}
    f_k^{d^*} (t) = \sqrt[4]{\frac{ Z_k(t) C^d_k(t)}{3 V \gamma_k} }\; \Biggr|_0^{f_{k, max}^d},
\label{eq:Freq_optim device}    
\end{equation}
where $W(\cdot)$ in (\ref{eq:Optim_rate}) denotes the principal branch of the Lambert function. 
\begin{figure}[t]
	\begin{center}
		   \includegraphics[width=8.8 cm]{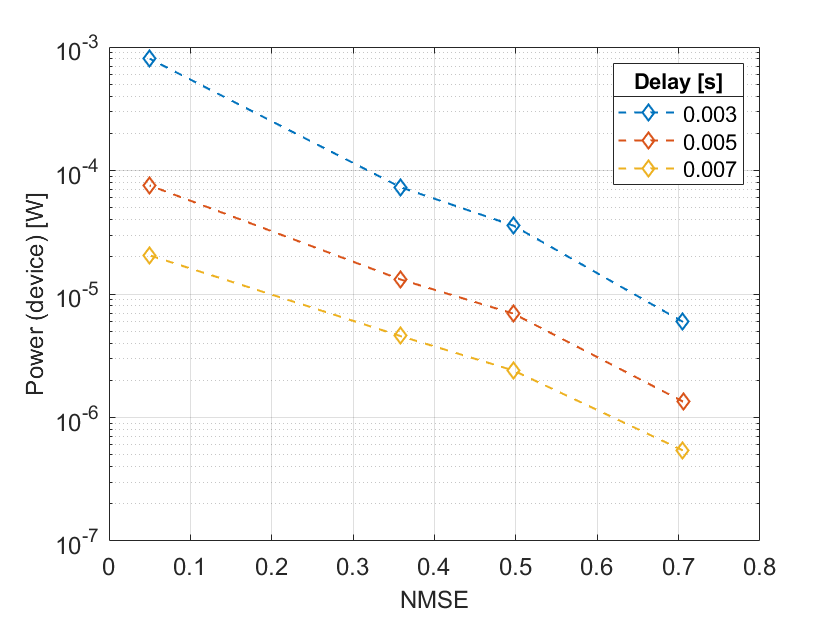}
	\end{center}
	\caption{Power-delay-accuracy trade-off for the GIB case.}
	\vspace{-15pt}
	\label{Tradeoff-device}
\end{figure}

The optimal $\beta^*_k(t) \in \mathcal{B}_k$ can then be found by simply searching the value in $\mathcal{B}_k$ that, together with (\ref{eq:Optim_rate}) and (\ref{eq:Freq_optim device}),  minimizes the objective of the sub-problem associated with device $k$. Finally, letting $C_k^s =\max_{\beta_k} C_{\beta_k}^s$, the optimal server frequency $f_c(t)$ and its split $\{f_k(t)\}_k$ among the devices are:
\begin{equation}
    f_c^* (t) = \frac{\sqrt{\sum_{k=1}^K \sqrt{Z_k(t) C^s_k}}}{\sqrt[4]{3V\eta}} \; \Biggr|_0^{f_{max}},
\label{eq:Freq_optim_tot}    
\end{equation} 
\begin{equation}
    f_k^* (t) = \frac{\sqrt{Z_k(t)C_k^s}}{\sqrt{\sum_{k=1}^K \sqrt{Z_k(t) C_k^s}}\sqrt[4]{3V\eta} },  \qquad \forall k.
\label{eq:Freq_optim}    
\end{equation}





\section{Numerical Results and Conclusion}
\vspace{-.1cm}

We start with the Gaussian case, where we take advantage of closed form expressions of the IB principle in (\ref{I(X;T)})-(\ref{I(T;Y)}), and of the MSE in (\ref{eq:MSE GIB}). \FP{We consider $K=100$ devices that send independent tasks to a common edge server. These devices are placed at a regular distance, from 5 to 150 meters, from the ES.  The maximum transmit power is $ p_{max} = 100 \; mW $. The access point operates with a carrier frequency $ f_0 = 1 \; GHz $. The wireless channels are generated using the Alpha-Beta-Gamma model from \cite {maccartney2016millimeter}. The bandwidth is set to $ B_k = 1 \; kHz $, and $ N_0 = -174 \; dBm/Hz $. Both server and devices are equipped with a $1.8 \; GHz $ CPU (Intel$^{\circledR}$ Celeron$^{\circledR}$  6305E Processor 4M Cache). In this scenario we have $ f_{max} = f_{k,max}^d = 1.8 \; GHz $, and $  \eta = \eta_k = 2.57*10 ^ {27} $, for all $k$. The input data $X$ has dimension $d_x=750$, whereas the output variable $Y$ has  $d_y=8$. Since there are only matrix multiplications, the number of operations to be performed at the device and at the ES are $C_k^d = d_x\textcolor{black}{d_t}$ and $C_{\beta_k}^s=\textcolor{black}{d_t}d_y$, where $d_t$ depends on $\beta_k$.}
\begin{figure}[t]
	\begin{center}
		   \includegraphics[width=8.80 cm]{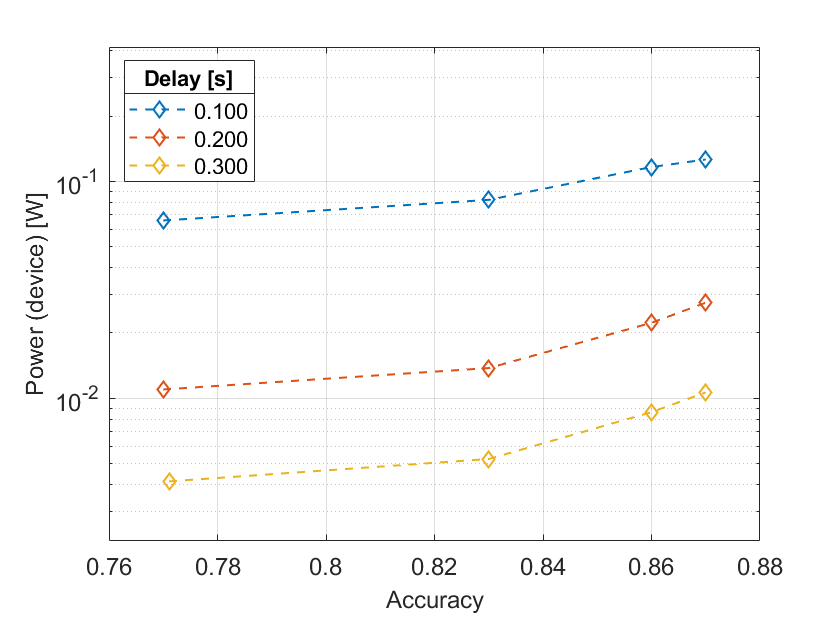}
	\end{center}
	\caption{Power-delay-accuracy trade-off for the CNN case.}
	\vspace{-15pt}
	\label{CNN-Tradeoff-device}
\end{figure}
In Fig. \ref{Tradeoff-device}, we illustrate the trade-off between average power consumption at the device side and average accuracy of the inference task, for different latency requirements. As expected, from Fig. \ref{Tradeoff-device}, we notice how a larger power consumption is needed to achieve a better learning accuracy, with a stricter latency constraint.

Then, we generalized the approach to a more practical case where the goal is image classification using a convolutional neural network (CNN). The idea is to split the neural network, executing the first layers at the device and the last layers at the edge server. The bottleneck is now represented by the amount of information available at each intermediate layers, which is a quantity that can be dynamically selected using our optimization method. 
The dataset contains $17000$ images equally distributed over $6$ different classes \cite{dataset} and it is split in $14000$ images for training and $3000$ for validation. The number of devices is $K=5$  and the bandwidth of each link is  $B_k=2$ MHz \FP{and $ f_0 = 6 \; GHz $. The number of operations at the device and at the ES is proportional to the number of operations required to compute the convolution of each layer.}
In Fig. \ref{CNN-Tradeoff-device}, we illustrate the trade-off between power, accuracy, and delay obtained by the proposed strategy and we can see that, also in this CNN case, a larger power is generally required to achieve a better accuracy  with a smaller delay requirement.

In this work we have proposed an edge learning scheme that combines the IB principle with stochastic optimization to dynamically identify and send only the information relevant to perform an inference task at an edge server on data collected by peripheral devices. The method does not require any prior knowledge of the wireless channel statistics and yields a dynamic allocation of radio and computation resources leading to a controllable trade-off between power consumption, learning accuracy and service delay. Further investigations are needed to exploit the IB principle to design a proper split of a neural network between source and destination sides.

\balance
\bibliographystyle{IEEEbib}
{\small
\bibliography{IB_biblio}}

\end{document}